\documentclass[aps,amsmath,amssymb,reprint,notitlepage]{revtex4-1}

\usepackage{color, enumitem, graphicx, hyperref, natbib}

\newcommand{\pe}{\text{Pe}}

\newcommand{\blu}[1]{#1}

\begin{document}

\title{Nonlinear dynamics of a chemically-active drop: from steady to chaotic self-propulsion}
\author{Matvey Morozov}
\affiliation{LadHyX -- D{\'e}partement de M{\'e}canique, 
  {\'E}cole Polytechnique -- CNRS, 91128 Palaiseau Cedex, France}
\author{S{\'e}bastien Michelin}
\email{sebastien.michelin@ladhyx.polytechnique.fr}
\affiliation{LadHyX -- D{\'e}partement de M{\'e}canique, 
  {\'E}cole Polytechnique -- CNRS, 91128 Palaiseau Cedex, France}

\begin{abstract}
Individual chemically active drops suspended in a surfactant solution were observed to self-propel spontaneously with straight, helical, or chaotic trajectories. To elucidate how these drops can exhibit such strikingly different dynamics and ``decide'' what to do, we propose a minimal axisymmetric model of a spherical active drop, and show that simple and linear interface properties can lead to both steady self-propulsion of the droplet as well as chaotic behavior. The model includes two different mobility mechanisms, namely, diffusiophoresis and the Marangoni effect, that convert self-generated gradients of surfactant concentration into the flow at the droplet surface. In turn, surface-driven flow initiates surfactant advection that is the only nonlinear mechanism and, thus, the only source of dynamical complexity in our model. Numerical investigation of the fully-coupled hydrodynamic and advection diffusion problems reveals that strong advection (e.g., large droplet size) may destabilize a steadily self-propelling drop; once destabilized, the droplet spontaneously stops and a symmetric extensile flow emerges. If advection is strengthened even further in comparison with molecular diffusion, the droplet may perform chaotic oscillations. Our results indicate that the thresholds of these instabilities depend heavily on the balance between diffusiophoresis and the Marangoni effect. Using linear stability analysis, we demonstrate that diffusiophoresis promotes the onset of high-order modes of monotonic instability of the motionless drop. We argue that diffusiophoresis has a similar effect on the instabilities of a moving drop.
\end{abstract}

\maketitle

\section{Introduction}
Self-propulsion of chemically-active systems has recently emerged as a canonical system of active colloids to study the behavior of active matter, where energy is introduced at the microcopic scale in the self-propulsion of individual agents~\cite{Marchetti16}. Among the many systems considered, catalytic (phoretic) rigid particles~\cite{Moran17} and chemically-active droplets~\cite{Maass16} have received a particular attention both experimentally and theoretically. Because of their small size, phoretic particles can be significantly influenced by Brownian fluctuations and a particular research focus on such systems can be found in their collective self-organization~\cite{Ginot18}.

In contrast, a fascinating feature of chemically-active droplets lies in their ability to exhibit complex dynamical behavior at the individual level as well.  Solitary active drops were observed to self-propel spontaneously with straight, helical, or chaotic trajectories, where the choice of a particular trajectory depends on the phase of the liquid crystal constituting the drop~\cite{Kruger16}, on the size of the droplet and on the intensity of the chemical reaction fueling the motion~\cite{Suga18}, as well as on the geometrical constraints~\cite{Jin18}. Self-deformation and division were shown to occur when drops are impregnated with surfactant~\cite{Caschera13}, so that active droplets were also recently considered as minimal model for synthetic cells~\cite{Nagasaka17}. At the collective level, and similarly to phoretic particles, active droplets can self-organize in complex clusters~\cite{Yang18b} in the presence of chemically-active species. Multiple active drops ``feel'' each other's presence and adjust their behavior: they may form ordered clusters~\cite{Thutupalli11, Weirich18}, repel~\cite{Moerman17}, or avoid crossing each other's trails~\cite{Kruger16}.

Experimental observations of active drops typically employ relatively small droplets with radius $\sim$100 $\mu$m or less. At those length scales, liquid drops are usually highly symmetric and, in the absence of external forcing, any kind of motion of a solitary active drop must initiate from a symmetry-breaking bifurcation~\cite{Herminghaus14, Yoshinaga17}. The properties of this bifurcation (or bifurcations) are yet not well understood. In particular, it is still unclear how multiple dynamical behaviors can arise for a single drop (e.g., straight or chaotic trajectories), and how a particular type of self-propelling mode is selected. The physico-chemical complexity of the different experimental systems considered (including the saturation of the droplet's surface by surfactant molecules or the nematic nature of the inner fluid) further leaves open several possible and potentially coupled origins for such complex dynamics. Instead of focusing on the detailed description of a particular experimental system, the present work aims to demonstrate, using a minimal yet generic model, that surface properties and in particular its mechanical response to self-generated physico-chemical gradients can provide the droplet with the ability  to exhibit both steady and chaotic self-propulsion.

Symmetry-breaking at the onset of drop self-propulsion originates from a self-induced  concentration gradient at the droplet interface of a chemical species, which is maintained despite diffusion by advective transport with the flow field generated by the droplet~\cite{Rednikov94,Izri14,Maass16}. For a general interface, and in contrast with strictly rigid particles, this flow field results from a mechanical forcing at the droplet's interface under the effect of the chemical gradient, through a combination of Marangoni effect and diffusiophoresis~\cite{Anderson89}. For fluid droplets, it is typically assumed that  Marangoni effect prevails, while diffusiophoresis is negligible~\cite{Izri14,Herminghaus14, Maass16}, due to the separation of scales between the drop's radius and the thickness of the interaction layer between the chemical species and the interface. This assumption is not always applicable, since some nanoparticle surfactants were observed to form a disordered, jammed assembly at the interface, thus rendering it immobile~\cite{Cui13}. In the case of an immobile interface (or large droplet viscosity), droplet can be considered as a particle and diffusiophoresis remains the only source of its mobility. In the present paper, we consider the general case including both diffusiophoresis and the Marangoni effect, and investigate how this dual behavior of the interface and the ratio of these two effects may influence the dynamics of the chemically active drop.

The paper is organized as follows. The minimal generic model for the self-propulsion of a chemically active drop is presented in Sec.~\ref{statement}. In Sec.~\ref{numerical} we outline and validate the methods of numerical analysis and the numerical results are presented in Sec.~\ref{nonlinear_regimes}. The main findings of the paper are finally discussed in Sec.~\ref{discussion}. 

\section{Problem statement}
\label{statement}

\subsection{Modelling active droplets}

We consider the dynamics of a force-free spherical droplet of radius $R$ suspended in the bulk of a surfactant solution. In recent experiments, active droplets with $R \sim 1$--$10 \, \mu$m were shown to spontaneously swim with velocities $U\sim 10$--$50 \, \mu$m.s$^{-1}$~\cite{Izri14, Moerman17, Suga18}, so that inertial forces in the fluid phases are negligible (i.e. the Reynolds number $\mbox{Re}=UR/\eta_o$ is exceedingly small, with $\eta_o$ the viscosity of the outer phase). The flow field, $\textbf{u}$, and pressure, $P$, therefore satisfy the equations of Stokes flow,
\blu{
\begin{align}
  \label{eqs_flow}
  \nabla \cdot \textbf{u}_i = 0, \quad
  & \nabla P_i = \eta_i \nabla^2 \textbf{u}_i, \\
  \label{eqs_flow2}
  \nabla \cdot \textbf{u}_o = 0, \quad
  & \nabla P_o = \eta_o \nabla^2 \textbf{u}_o,
\end{align}}
with subscripts $i$ and $o$ denoting the corresponding quantity inside and outside of the droplet, respectively. Assuming that surfactant molecules do not penetrate into the droplet, the concentration of surfactant molecules $C$ outside the drop satisfy the following advection-diffusion \blu{equation},
\begin{equation}
  \label{eqs_ad}
  \frac{\partial C}{\partial t} + \textbf{u}_o \cdot \nabla C 
    = {\cal D} \nabla^2 C,
\end{equation}
where ${\cal D}$ is the molecular diffusivity of the surfactant.

The physico-chemical activity of swimming droplets can involve several mechanisms, including  micellar and molecular pathways to the droplet dissolution~\cite{Herminghaus14,Izri14, Moerman17, Suga18}. The former involves the dissolution of the droplet by micelles present in the surfactant-saturated outer phase~\cite{Izri14}, while in the latter, droplet dissolution is achieved through the formation of swollen micelles from the surfactant molecules present in the outer phase~\cite{Moerman17}. In the following, we specifically consider the molecular pathway, although the formalism presented here could easily be extended to account for other dissolution mechanisms. In this framework, the drop undergoes gradual dissolution sustained by a chemical reaction at the fluid-fluid interface. In the simplest possible case, the reaction rate is fixed and the drop consumes surfactant molecules at a fixed rate ${\cal A} > 0$,
\begin{equation}
  \label{c_consump}
  {\cal D} \,\mathbf{n}\cdot\nabla C= {\cal A}
  \quad \text{at  } r = R.
\end{equation}

The mobility of the drop arises from inhomogeneity in surfactant concentration and, in general, may come from two distinct interfacial mechanisms. The first is diffusiophoresis, taken into account by a nonzero slip velocity at the droplet interface,
\begin{equation}
  \label{slip}
  \textbf{u}_i - \textbf{u}_o
    = {\cal M} \left( \mathbf{I} - \textbf{n}\textbf{n} \right) \cdot\nabla C
  \quad \text{at  } r = R,
\end{equation}
where ${\cal M}$ is the mobility coefficient, $\textbf{n}$ is the outward normal to the droplet surface, $\mathbf{I}$ is the identity tensor. Since surfactant molecules are attracted to the interface during micellar dissolution, \blu{we assume that any repulsive interactions between the droplet interface and the surfactant molecules are negligible and postulate that ${{\cal M} \geq 0}$~\cite{Anderson89}.} The second mechanism is the Marangoni effect, stemming from uneven surface tension at the fluid-fluid interface. In particular, we assume that surface tension depends linearly on the surfactant concentration at the interface,
\begin{equation}
  \label{gamma_c}
  \gamma = \gamma_0 - \gamma_C \left( C - C_\infty + {\cal A} R \right),
\end{equation}
where $\gamma_0$ denotes the reference value of surface tension measured at $C = C_\infty - {\cal A} R$, whereas $\gamma_C$ is a positive constant. Note that our initial assumption of a spherical droplet requires capillary pressures to dominate hydrodynamic stresses, so that the typical capillary number, $\mbox{Ca}=\eta_oU/\gamma_0$ is very small -- see Ref.~\cite{Morozov18} for a generalization of this framework to deformable droplets.

\begin{figure}
\centering
  \includegraphics[scale=0.53]{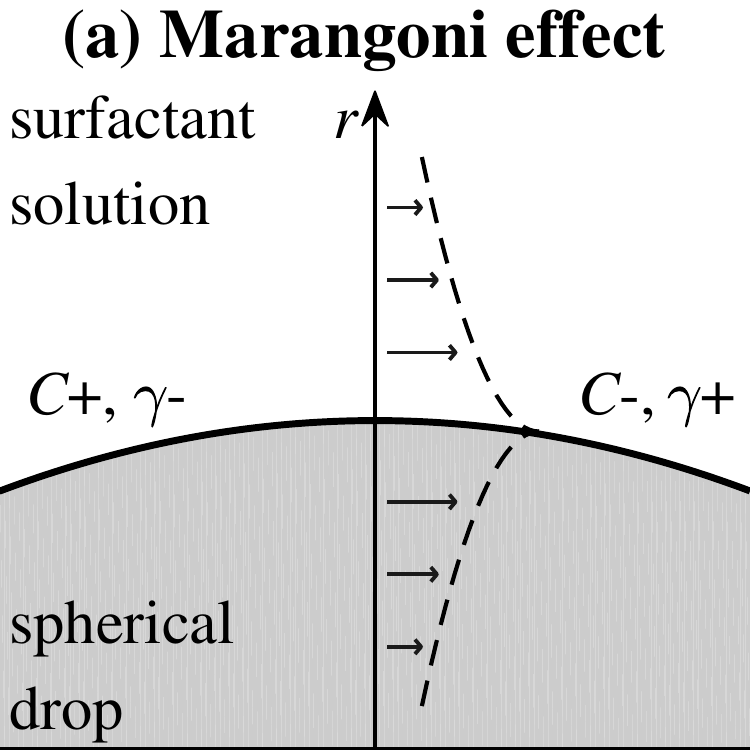}
  \quad
  \includegraphics[scale=0.53]{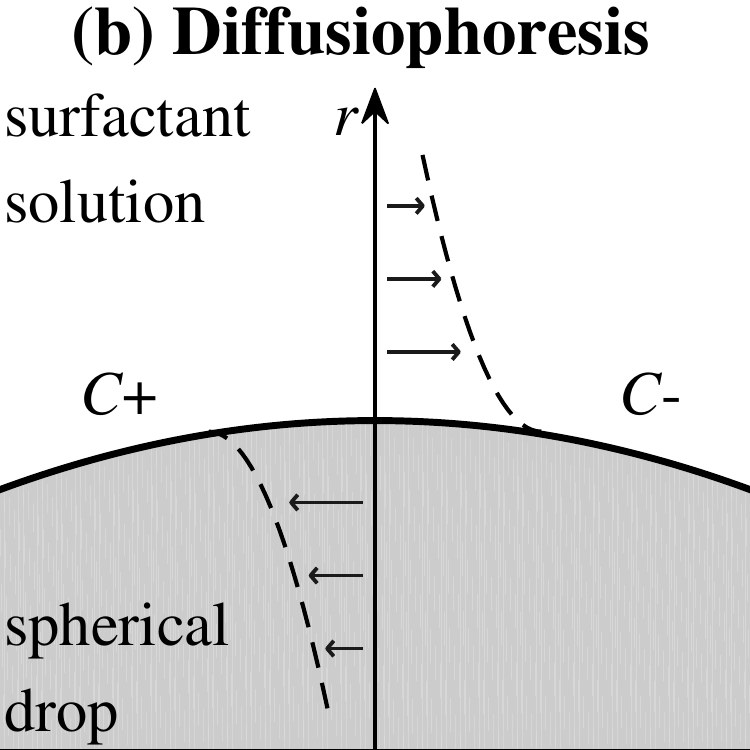}
  \caption{
    Sketch of the flow field established by each  mobility mechanism in response to a surfactant concentration disturbance, $C+ > C-$. The flow is shown in the reference frame of a quiescent drop.
    (a) Marangoni effect:
    a concentration disturbance results in uneven surface tension, 
    $\gamma+ > \gamma-$ and discontinuous tangential stresses, while the flow velocity remains continuous at the interface.
    (b) Diffusiophoresis:
    hydrodynamic stresses are continuous (no interfacial stress), but a discontinuity in flow velocity arises from the concentration contrast. 
    Note that direction of the flow within the drop depends on the 
    dominating interfacial mechanism, as demonstrated in Eq.~\ref{ain_sol}.
  }
  \label{flow_sketch}
\end{figure}

Uneven surface tension contributes to the balance \blu{of} stresses at the interface. In the limit of a nondeformable droplet, it is sufficient to only consider the balance of tangential stresses,
\begin{equation}
  \label{stress}
  \textbf{n} \cdot \left( \boldsymbol \sigma_i - \boldsymbol \sigma_o \right)
    = -\gamma_C \left( I - \textbf{n}\textbf{n} \right) \nabla C
  \quad \text{at  } r = R,
\end{equation}
where $\boldsymbol \sigma_i$ and $\boldsymbol \sigma_o$ denote the stress tensor of the fluid within and outside of the drop. Fundamental differences between the two mobility mechanisms included in the model are highlighted in Fig.~\ref{flow_sketch}: mobility due to the Marangoni effect is characterized by a continuous velocity field and discontinuous interfacial stresses, while diffusiophoresis results in discontinuous velocity field and continuous interfacial stresses.

Far away from the droplet, the flow velocity in the frame of reference of the droplet, and the surfactant concentration attain constant values,
\begin{equation}
  \label{far}
  \textbf{u}_o = -{\cal U}_\infty \textbf{e}_z, \quad 
  C = {\cal C}_\infty,
\end{equation}
where $\textbf{e}_z$ is the unit vector directed along the symmetry axis of the problem and ${\cal U}_\infty \textbf{e}_z$ corresponds to the droplet self-propulsion velocity determined from the condition that the total hydrodynamic force on the droplet vanishes,
\begin{equation}
  \label{force}
  \int_{r=R} \boldsymbol \sigma_o \cdot \mathbf{n} \mathrm{d}S=0.
\end{equation}

\blu{It is easy to see that in the limit of $\eta_i \rightarrow \infty$, the Stokes equation within the drop reads ${\nabla^2 \textbf{u}_i = 0}$, while the balance of stresses reduces to ${\textbf{n} \cdot \boldsymbol \sigma_i = 0}$. Naturally, ${\textbf{u}_i = 0}$ in this limit and the problem statement becomes identical to the model considered by Michelin~{\it et al.}~\cite{Michelin13b}.}

\subsection{Nondimensionalization}
In what follows, all quantities are nondimensionalized, using the droplet's radius $R$ as characteristic length scale, and scaling the relative concentration of surfactant (i.e., $C - {\cal C}_\infty$) by ${\cal A} R / {\cal D}$. We further define the velocity scale as the terminal velocity of the droplet moving in a surfactant gradient ${{\cal A}/{\cal D}}$ due to both diffusiophoresis and Marangoni effect~\cite{Anderson89},
\begin{equation}
  \label{u_scale}
  {\cal V} \equiv 
    \frac{{\cal A} \left( \gamma_C R + 3 \eta_i {\cal M} \right)}
      {{\cal D} \left( 2 \eta_o + 3 \eta_i \right)}.
\end{equation}
Finally, the characteristic time-scale is chosen as $R / {\cal V}$.

Dimensionless form of Eqs.~\eqref{eqs_flow}-\eqref{slip} and~\eqref{stress} includes three dimensionless parameters,
\begin{equation}
\pe \equiv \frac{{\cal V} R }{ {\cal D}},\quad \eta \equiv \frac{\eta_i }{ \eta_o},\quad m \equiv \frac{\eta_i {\cal M} }{ \gamma_C R},
\end{equation} 
which are respectively (i) the P{\'e}clet number, ${\pe}$, which measures the relative influence of advection and diffusion in surfactant transport but can also be seen as a measure of the droplet's size, (ii) the viscosity contrast ${\eta}$ between the inner and outer phases, and (iii) the mobility contrast ${m\geq 0}$ which compares the terminal velocity of the drop driven exclusively by diffusiophoresis to its counterpart achieved in response to the Marangoni effect. Therefore, ${m = 0}$ corresponds to the motion driven purely by the Marangoni effect, while the drop self-propelling by diffusiophoresis only features ${m \rightarrow \infty}$.

\subsection{Axisymmetric Stokes flow}
\label{stokes_flow}
In the following, we focus on a single droplet in an infinite fluid domain. For simplicity, we thus assume that the flow field within and around the spherical drop is axisymmetric and, thus, can be expressed in axisymmetric spherical coordinates in terms of a stream function $\psi_{i,o}(t,r,\mu)$ with $\mu = \cos \theta$. In this case, the general solution of the Stokes equations~\eqref{eqs_flow}-\eqref{eqs_flow2} is given by the Lamb solution~\citep{Lamb45, Happel83, Leal07} with the flow outside the droplet converging to a finite unidirectional flow as $r \rightarrow \infty$, while the flow within the drop is regular at the origin, namely,
\begin{align}
  \label{innerflow}
  \psi_i \left( t, r, \mu \right)
  & = \sum\limits_{n=1}^\infty 
      a_{i,n}(t) r^{n+1} \left( 1 - r^2 \right)
        \left( 1 - \mu^2 \right) L_n'(\mu), \\
  \label{outerflow}
  \psi_o \left( t, r, \mu \right)
  & = \sum\limits_{n=1}^\infty 
      a_{o,n}(t) \Psi_n(r) \left( 1 - \mu^2 \right) L_n'(\mu), \\
  \textrm{with   }\Psi_n \left( r \right) 
  & = \begin{cases}
    \displaystyle\frac{1}{r} - r^2, &n = 1 \\[8pt]
    \displaystyle\frac{1 - r^2}{r^n}, &n > 1
  \end{cases},
\end{align}
where $L_n(\mu)$ is the $n$-th Legendre polynomial, prime denotes the derivative, and $a_{i,n}(t)$ and $a_{o,n}(t)$ are unknown functions of time. Naturally, the Stokeslet term is omitted in~\eqref{outerflow} since the droplet is force-free~\cite{Blake70}. Also note that equations~\eqref{innerflow}-\eqref{outerflow} imply that the flow velocity decreases away from the interface both within and outside of the drop, as expected since the fluid and droplet motions results from an interfacial forcing.

\section{Numerical modeling of the nonlinear dynamics}
\label{numerical}
\subsection{Presentation of the numerical method}
In this section, we present the numerical methods used to solve jointly for the hydrodynamic problem and surfactant dynamics. Following Michelin and Lauga, we expand the surfactant distribution around the droplet as a truncated series of Legendre harmonics~\cite{Michelin13a},
\begin{equation}
  \label{c_num_exp}
  C \left( t, r, \mu \right) 
    = \sum\limits_{n=0}^N C_n \left( t, r \right) L_n \left( \mu \right),
\end{equation}
with $N$ sufficiently large so as to ensure proper convergence of the description of the surfactant dynamics. Substitution of approximation~\eqref{c_num_exp} along with expansions~\eqref{innerflow}--\eqref{outerflow} into the dimensionless form of the boundary conditions~\eqref{slip} and~\eqref{stress} and subsequent projection of the result onto the $n$-th Legendre polynomial provides a direct one-to-one relationship at each order $n$ between the amplitudes of the hydrodynamic modes, $a_{i,n}(t)$ and $a_{o,n}(t)$, and the value of the concentration modes at the drop's surface, $C_n \left( t, 1 \right)$,
\begin{equation}
  \label{ai_ao_sol}
  a_{i,n}(t) = A_{i,n} C_n \left( t, 1 \right), \quad
  a_{o,n}(t) = A_{o,n} C_n \left( t, 1 \right),
\end{equation}
where the transfer coefficients $A_{i,n}$ and $A_{o,n}$ are given by
\begin{align}
  \label{ain_sol}
  & A_{i,n} = \begin{cases}
    \dfrac{\eta - 2 m}{2 \eta \left( 1 + 3 m \right)}, & n = 1\\[8pt]
    \dfrac{ \left( \eta - m \left[ 2 n + 1 \right] \right) 
      \left( 2 + 3 \eta \right)}
    {2 \eta \left( 2 n + 1 \right) \left( 1 + \eta \right) 
      \left( 1 + 3 m \right) },
    & n > 1
  \end{cases}, \\
  \label{aon_sol}
  & A_{o,n} = \begin{cases}
    1/3, & n = 1\\[3pt]
    \dfrac{ \left( 1 + m \left[ 2 n + 1 \right] \right) 
      \left( 2 + 3 \eta \right)}
    {2 \left( 2 n + 1 \right) \left( 1 + \eta \right) 
      \left( 1 + 3 m \right) },
    & n > 1
  \end{cases}.
\end{align}
Equations~\eqref{ain_sol} and~\eqref{aon_sol} display two important features. First, for ${m > \eta / 2}$, all coefficients $A_{i,n}$ become negative and the flow direction within the drop is reversed, as illustrated in Figs.~\ref{flow_sketch}a and~\ref{flow_sketch}b. Such flow reversal is a typical feature of the drops propelled by phoretic effects~\cite{Yang18}.

Second, in contrast to the pure Marangoni case (where ${m=0}$ and ${A_{i,n}, A_{o,n}} \rightarrow 0$ as $n \rightarrow \infty$), there is no natural ``damping'' of higher-order Legendre modes in the presence of diffusiophoresis: for ${m \neq 0}$, transfer coefficients $A_{i,n}$ and $A_{o,n}$ remain finite as $n \rightarrow \infty$.  In other words, the amplitude of higher-order modes typically increases with $m$. In appendix~\ref{asymptotic}, we elucidate this effect by means of linear stability analysis and demonstrate that monotonic instability thresholds decrease with $m$ as shown in Fig.~\ref{thresholds}. Note that the hydrodynamic and concentration mode amplitudes $a_{i,n}$, $a_{o,n}$ and $C_n$ still asymptotically decay for $n\rightarrow\infty$, ensuring the convergence of the expansion in Eq.~\eqref{c_num_exp}.

We substitute solution~\eqref{ai_ao_sol} into the projection of the dimensionless form of the advection-diffusion equation~\eqref{eqs_ad} onto the $n$-th Legendre polynomial and obtain a set of $N$ coupled differential equations describing the evolution of $C_n \left( t, r \right)$, namely,
\begin{widetext}
\begin{multline}
\label{cn_evo}
  \frac{\partial C_n}{\partial t}
  = \pe^{-1} \left( \frac{\partial ^2C_n}{\partial r^2} + \frac{2}{r} \frac{\partial C_n}{\partial r} 
    - \frac{n \left( n + 1 \right) C_n}{r^2} \right)
  \\ - \frac{2 n + 1}{2 r^2} \sum\limits_{j=1}^N \sum\limits_{k=0}^N 
    A_{o,j} \left. C_j \right|_{r=1}
    \Big[ j \left( j + 1 \right) I_{jkn} \Psi_j \frac{\partial C_k}{\partial r}
      + J_{jkn} \frac{\mathrm{d}\Psi_j}{\mathrm{d} r} C_k \Big],
\end{multline}
\end{widetext}
with
\blu{
\begin{align}
  & I_{jkn} \equiv \int\limits_{-1}^1 L_j(\mu) L_k(\mu) L_n(\mu) d\mu, \\
  & J_{jkn} \equiv \int\limits_{-1}^1
    \left( 1 - \mu^2 \right) L_j'(\mu) L_k'(\mu) L_n(\mu) d\mu.
\end{align}
}
Boundary conditions for Eq.~\eqref{cn_evo} are given by projection of Eqs.~\eqref{c_consump},~\eqref{far} onto the basis of Legendre polynomials,
\begin{equation}
\label{cn_evo_bc}
  \left. \frac{\partial C_n}{\partial r} \right|_{r = 1} = \begin{cases}
    1, & n = 0\\
    0, & n > 0
  \end{cases}, \qquad
  \left. C_n \right|_{r \rightarrow \infty} = 0.
\end{equation}

Similarly to Refs.~\cite{Michelin13a,Michelin14}, we solve the set of evolution equations~\eqref{cn_evo} numerically, using an explicit time-stepping scheme for the advective term and the Crank-Nicholson scheme for the diffusive term. We employ an exponentially stretched spatial grid, namely, $r = e^{\xi^3 - 1}$, where $\xi$ is evenly spaced. In our computations we use spatial grids with $60$ and $120$ nodes and the time step of $0.05$ and $0.02$, respectively. We further ensure the convergence of the modal approximation~\eqref{c_num_exp} by repeating all of the computations for $N = 30$, $35$, and $40$ modes.

\subsection{Validation of the numerical method}
\label{validation}
The numerical method presented above is first validated against the predictions of the asymptotic analysis for the onset of self-propulsion and saturated velocity near the threshold, as carried out in Appendix~\ref{asymptotic}. In particular, this analysis reveals that (i) ${\pe_1 = 4}$ is the instability threshold corresponding to the onset of spontaneous self-propulsion and (ii) in vicinity of the threshold self-propulsion velocity is ${U_\infty = (\pe - \pe_1 ) / 16}$ (see also Ref.~\cite{Morozov18}).

\begin{figure*}
\centering
  \raisebox{1.75in}{(a)}
  \includegraphics[scale=0.54]{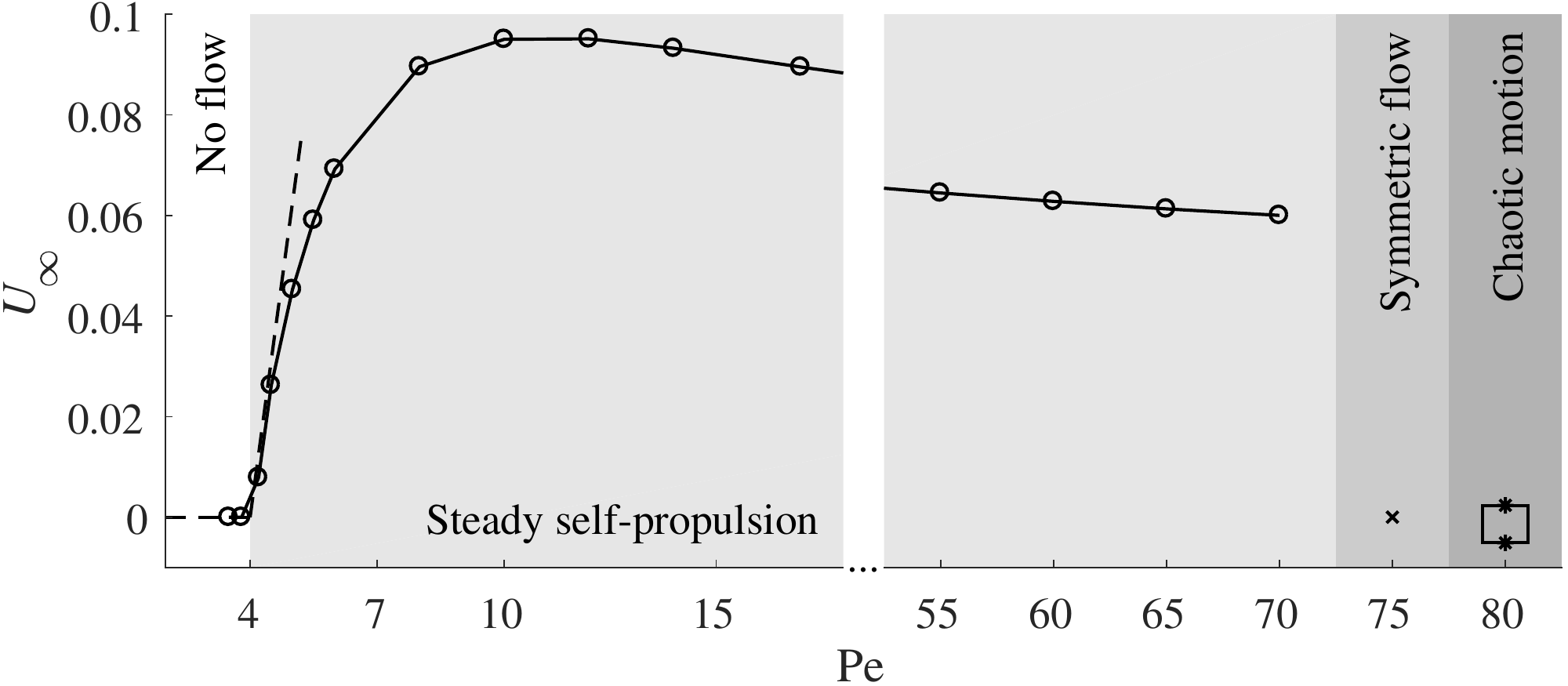}
  \quad
  \raisebox{1.75in}{(b)}
  \includegraphics[scale=0.54]{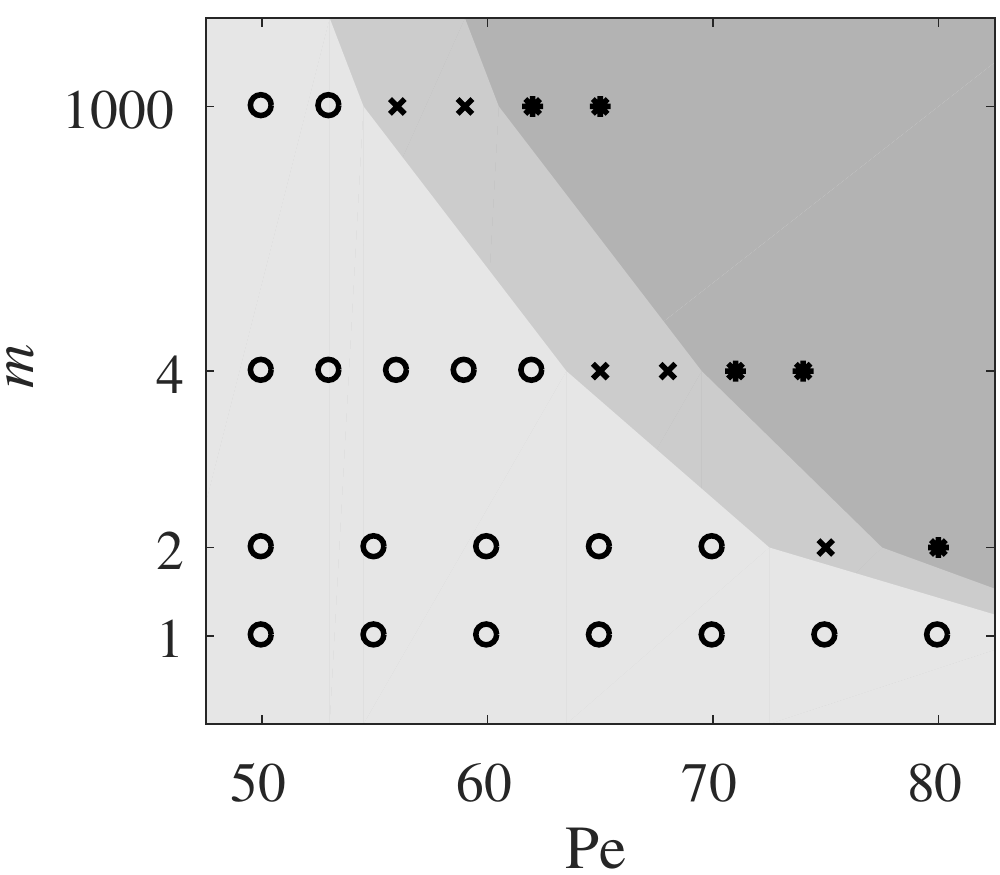}
  \caption{
    (a) Evolution of the self-propulsion velocity $U_\infty$
    with the P{\'e}clet number for ${m = 2}$ and ${\eta=1}$.
    Dashed line represents the result of asymptotic analysis,
    ${U_\infty = ( \pe - 4 ) / 16}$.
    For chaotic oscillations (${\pe \geq 80}$), the range of velocities is shown.
    (b) Dynamical regime observed in the computations for varying P\'eclet number $\pe$ and mobility ratio $m$:
    steady self-propulsion ($\circ$),
    steady symmetric extensile flow ($\times$),
    and chaotic oscillations ($*$).
  }
  \label{regimes}
\end{figure*}

These two findings are \blu{compared} to the results of the nonlinear numerical simulations as follows. Setting ${m=2}$ and ${\eta=1}$, for a discrete set of values of $\pe$ ($\pe=3.5$, $3.8$, $4.2$, $4.5$, $5$, $5.5$, and $6$), the numerical simulation is initiated by adding to the isotropic steady state~\eqref{base} a small asymmetric perturbation. For $\pe>4$, after a transient characterized by an exponential growth of the swimming velocity, a new anisotropic steady state is reached, and the terminal velocity of the drop in these computations agrees well with the theoretical predictions, as shown in the left part of Fig.~\ref{regimes}a.

\section{Nonlinear dynamics of an active droplet}
\label{nonlinear_regimes}
The dimensionless form of Eqs.~\eqref{eqs_flow}-\eqref{force} describes the joint dynamics of the surfactant concentration and flow fields, and allows for a trivial isotropic solution where the droplet is stationary and no fluid motion arises as the concentration distribution is isotropic. This isotropic state loses stability when advection of the surfactant concentration is sufficiently large, i.e. beyond a critical $\pe$ (see Appendix~\ref{asymptotic}, and Refs.~\cite{Michelin13b,Izri14}).

The main goal of the present work, and central purpose of this section, is to investigate the droplet dynamics away from the instability threshold. To this end, we perform the computations with ${m=2}$, ${\eta=1}$, and sequentially increasing values of the P{\'e}clet number, where each computation employs the limit regime achieved at the previous value of $\pe$ as an initial condition (steady solution obtained in Sec.~\ref{validation} for $\pe = 6$ is used to initialize the first computation). This continuation procedure yields a set of dynamical regimes which we discuss below.

\subsection{Steady self-propulsion}
For $\pe\geq 4$ and up to $\pe = 70$, the long-time dynamics is that of a steadily self-propelling drop (i.e. $\mathcal{U}_\infty\neq 0$). Similarly to the numerical results of Refs.~\cite{Michelin13b,Izri14}, droplet self-propulsion velocity reaches a maximum value around $\pe = 10$ and then decreases gradually with increasing $\pe$. Decrease in self-propulsion velocity suggests that strong advection hinders the formation of the concentration gradient propelling the drop, a feature that was already identified in the propulsion of chemically-asymmetric particles at finite and large $\pe$~\cite{Julicher09,Michelin14,Yariv15}.

\begin{figure}
  \centering
  \includegraphics[scale=0.75]{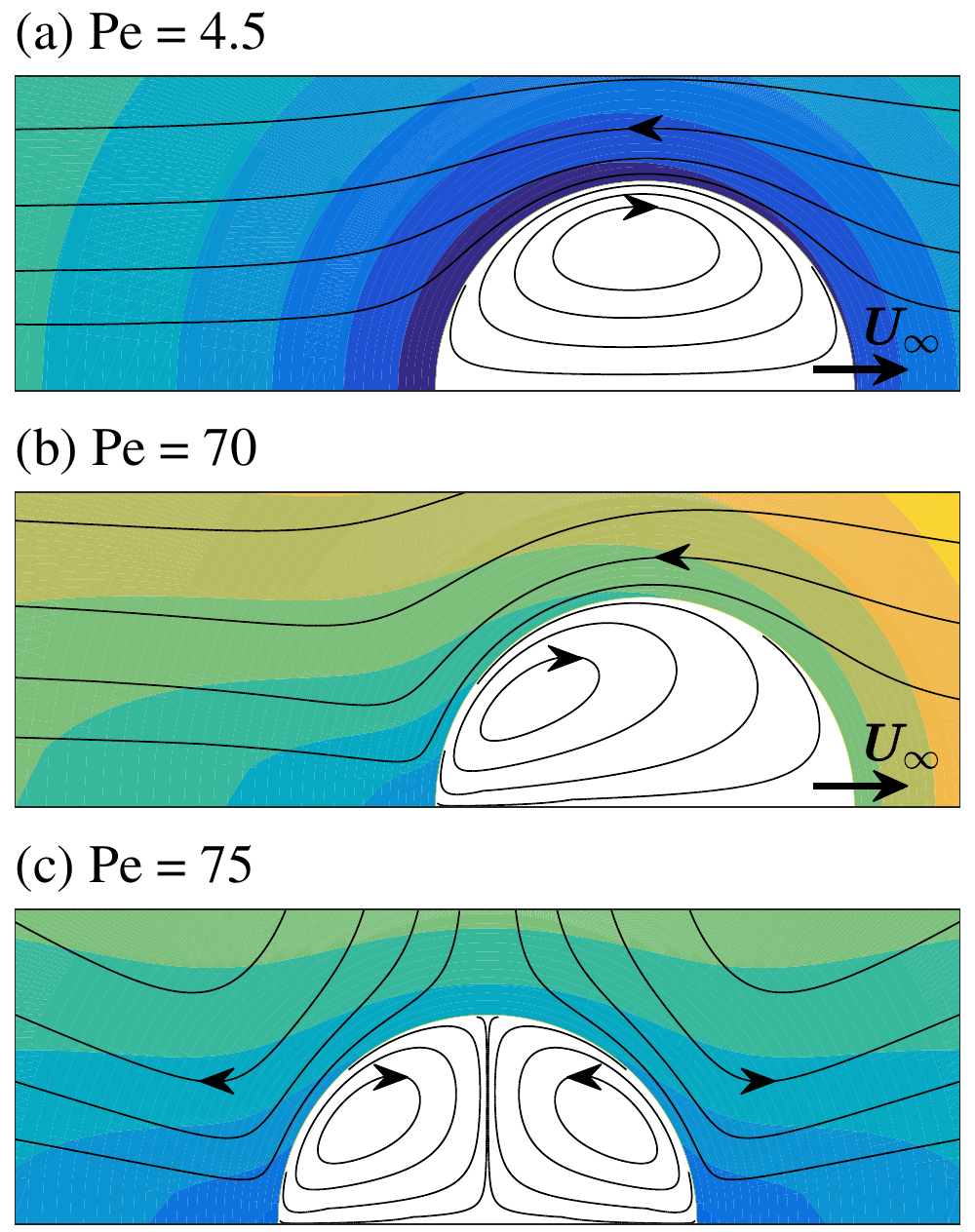}
  \caption{
    Concentration distribution around the drop (color map)
    and stream lines (lines and arrows)
    \blu{for ${m=2}$, ${\eta=1}$ and increasing P{\'e}clet number:
    (a) ${\pe = 4.5}$, (b) ${\pe = 70}$ -- drop self-propelling to the right,
    and (c) ${\pe = 75}$ -- stationary drop stirring a symmetric flow.
    Flow field is shown in the reference frame of the drop.
    In this paper, 
    the flow and concentration fields are assumed axisymmetric,
    thus, only half of the spherical drop is shown.
    Vertical axis corresponds to the distance from the axis of symmetry.}
  }
  \label{flows1}
\end{figure}

In figures~\ref{flows1}a and~\ref{flows1}b, we demonstrate that at high P{\'e}clet number, the surfactant concentration at the droplet surface resulting from the advection-diffusion dynamics is almost uniform at the front of the propelling drop, while the rear surface of the droplet experiences larger concentration gradients, and thus stronger mechanical forcing: as a results, the recirculation vortex within the drop is pushed towards its back as $\pe$ increases.

This can be further understood as follows. The flow velocity outside the droplet is characterized by two stagnation points in the front and at the back of the droplet. Near the rear stagnation point, the flow leaves the droplet and for large $\pe$, the significant advection of the surfactant results in an enhanced surfactant-depleted wake. In constrast, near the front of the droplet, advection of surfactant-rich fluid toward the droplet's surface maintains a rather large and uniform concentration level. 

\begin{figure}
\centering
  \includegraphics[scale=0.75]{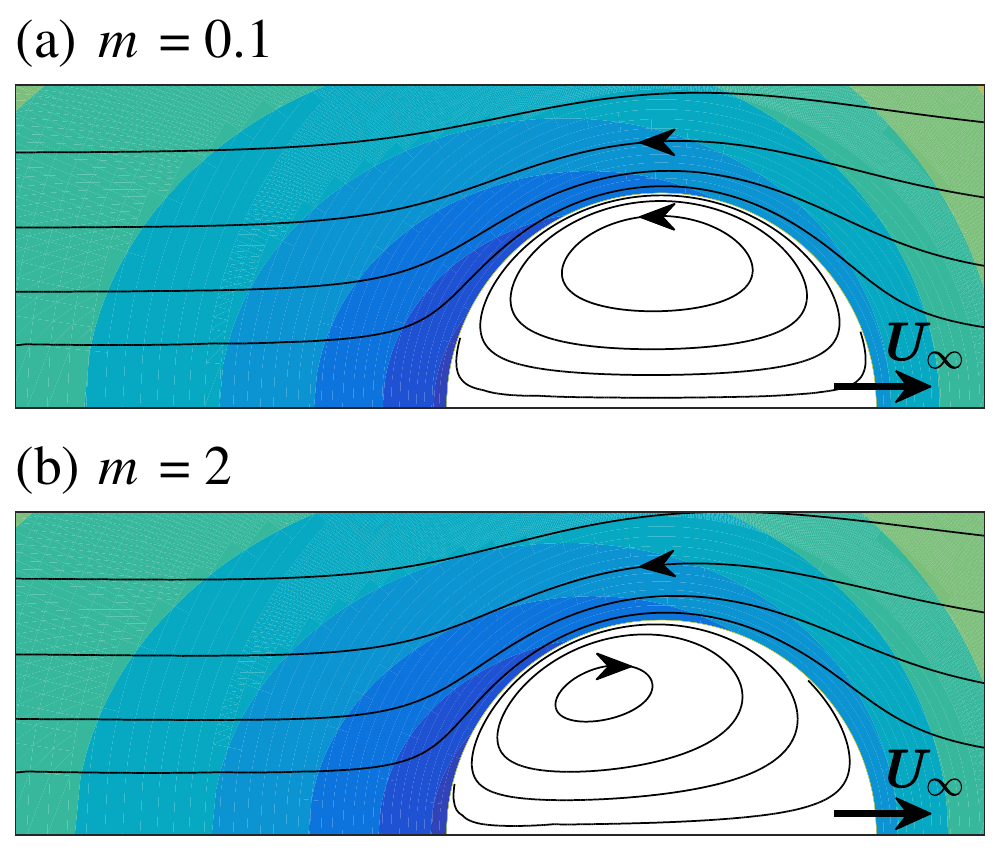}
  \caption{
    Concentration distribution around the drop (color map)
    and stream lines (lines and arrows)
    \blu{for ${\pe=8}$, ${\eta=1}$ and increasing mobility contrast $m$.
    Flow field is shown in the reference frame of the drop.
    Drop self-propels to the right in both panels.
    In this paper, 
    the flow and concentration fields are assumed axisymmetric,
    thus, only half of the spherical drop is shown.
    Vertical axis corresponds to the distance from the axis of symmetry.}
  }
  \label{flows2}
\end{figure}

We further conduct additional computations with ${\pe = 8}$, ${m = 0.1}$, and ${\eta = 1}$ in order to demonstrate that assymetry of the flow within the drop depends on the value of mobility contrast. As we argued in the discussion of Eqs.~\eqref{ain_sol} and~\eqref{aon_sol}, the values of the higher-order transfer coefficients $A_{i,n}$ and $A_{o,n}$ increase with $m$. Naturally, an increase of the transfer coefficients results in an enhanced flow field for a given concentration distribution and, thus, enhanced advection by higher-order azimuthal modes. As a result, the flow field observed at ${m = 0.1}$ (Fig.~\ref{flows2}a) appears more symmetric than its counterpart obtained at ${m=2}$ (Fig.~\ref{flows2}b), for which a larger signature of the higher order modes is observed in the focusing of the recirculation zone at the back of the droplet. Figure~\ref{flows2} also demonstrates the reversal in the direction of flow circulation within the droplet when $m$ is increased: in a Marangoni-dominated regime ($m=0.1$, Fig.~\ref{flows2}a), the flow velocity on both sides of the interface is oriented toward the back of the droplet and a Marangoni stress is exerted from the back of the droplet. When diffusiophoresis becomes significant ($m=2$, Fig.~\ref{flows2}b), the discontinuity of the flow velocity at the surface arising from the surface concentration gradient becomes large enough to drive the flow within the droplet in the opposite direction (see also Fig.~\ref{flow_sketch}).

\subsection{Steady symmetric extensile flow}
Using the continuation method, the results of our computations at ${\pe=75}$ are in stark contrast to the self-propelled state described above and instead result in a steady symmetric extensile flow with the concentration distribution shown in Fig.~\ref{flows1}c. We argue that the steady self-propulsion regime becomes unstable at this point due to the nonlinear advective coupling, and the system reaches a different branch of steady states characterized by no net propulsion and a dominance of the $n=2$ azimuthal mode. Specifically, strong advection skews surfactant distribution around the drop: surfactant concentration at the front part of the drop becomes almost constant, while a small depleted zone is pushed towards the back. In turn, region of constant surfactant concentration is associated with locally weakened interfacial flow that becomes unstable with respect to higher-order, symmetric modes of instability.

The threshold (i.e., critical $\pe$) for spontaneous flow symmetrization further depends on the value of the mobility contrast $m$, as demonstrated by repeating this analysis for ${m=1}$, ${m=4}$, and ${m=1000}$, using the continuation procedure described in Sec.~\ref{nonlinear_regimes}. Our results, summarized in Fig.~\ref{regimes}b, indeed indicate that the spontaneous symmetrization threshold decreases with $m$, however, the rate of the decrease is reduced drastically when ${m \gg 1}$. Based on these results, we hypothesize that spontaneous flow symmetrization relies on the higher-order terms of the modal expansion~\eqref{outerflow} which are effectively damped when $m < 1$, as argued in the discussion of Eqs.~\eqref{ain_sol}--\eqref{aon_sol}. In Fig.~\ref{regimes}b, results are presented for P{\'e}clet number smaller than  ${\pe=80}$; beyond ${\pe=80}$, convergence of the expansion in Eq.~\eqref{c_num_exp} requires an increase in the number of azimuthal modes considered, and as a result the computational cost is sharply increased in that region.

\subsection{Chaotic oscillations}
\begin{figure*}
\centering
  \raisebox{1.35in}{(a)}
  \includegraphics[scale=0.5]{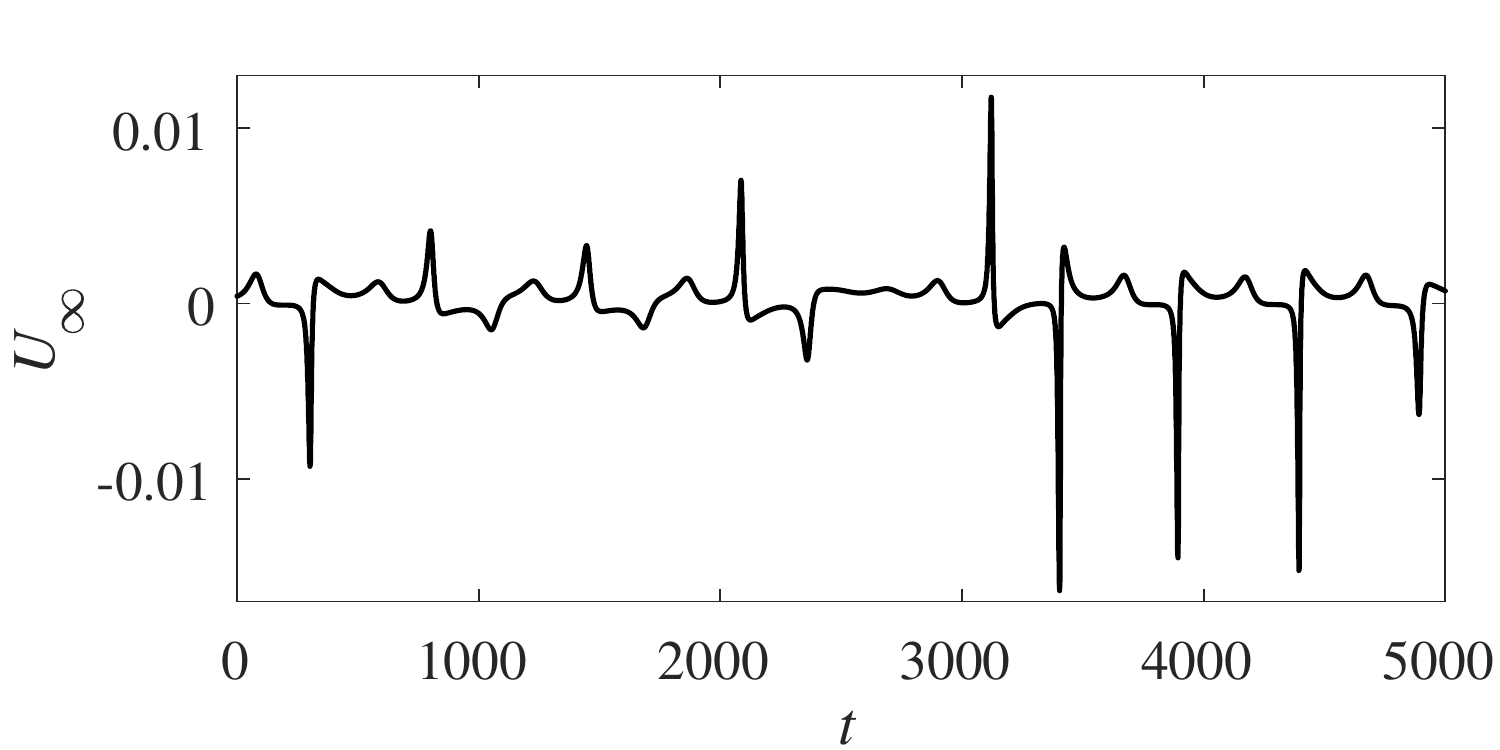}
  \quad
  \raisebox{1.35in}{(b)}
  \includegraphics[scale=0.5]{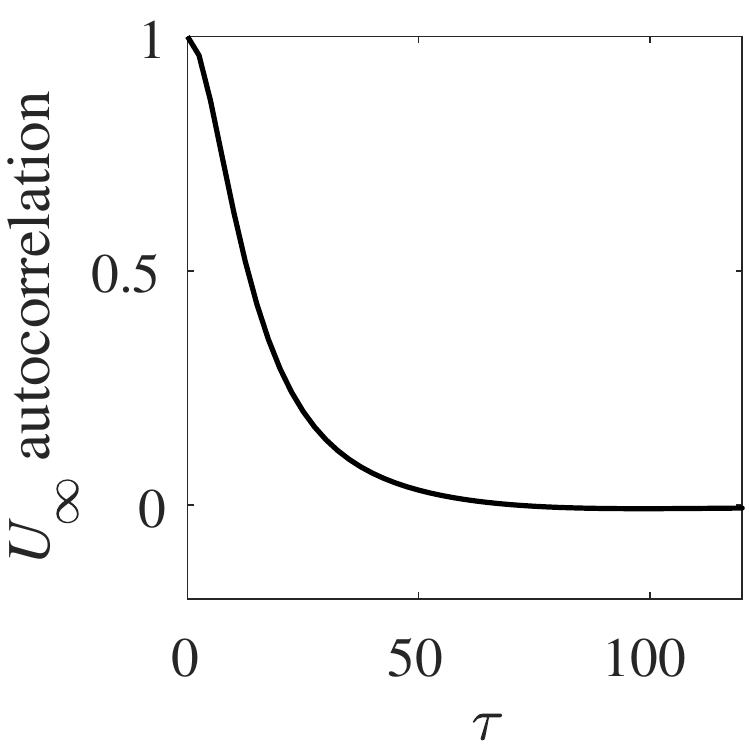}
  \quad
  \raisebox{1.35in}{(c)}
  \includegraphics[scale=0.5]{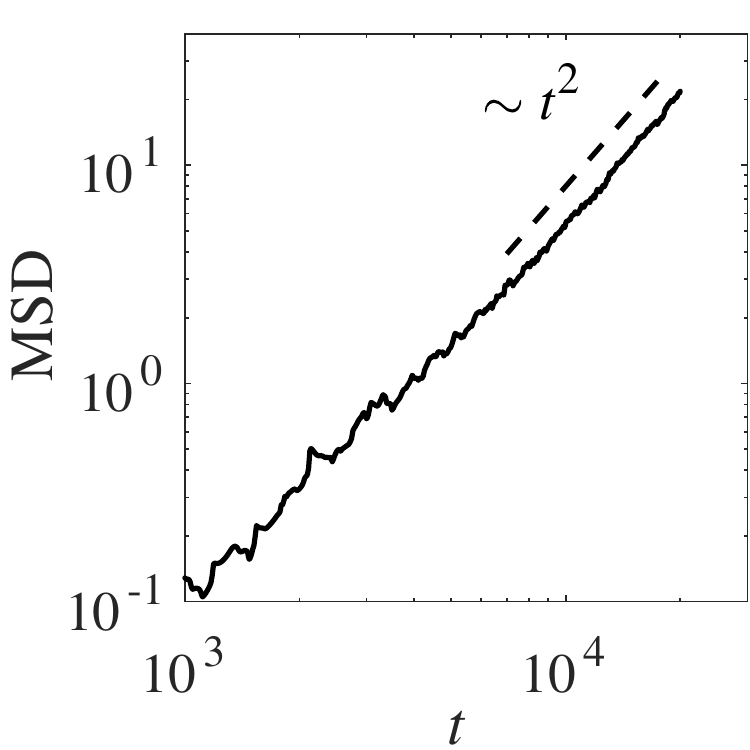}
  \caption{
    Chaotic oscillations observed beyond the threshold of spontaneous symmetrization
    at ${\pe=80}$ for ${m=2}$ and ${\eta=1}$.
    (a) Typical unsteady evolution of the drop velocity.
    (b) Autocorrelation of the drop velocity.
    (c) Mean square displacement (MSD) of the drop performing chaotic oscillations.
  }
  \label{chaos}
\end{figure*}

For $\eta=1$ and ${m \geq 2}$, increase of the P{\'e}clet number beyond the spontaneous symmetrization threshold results in the onset of chaotic oscillations illustrated in Fig.~\ref{chaos}. The oscillations are characterized by short intervals of larger self-propulsion velocity in random directions. This demonstrates that the interplay of diffusiophoresis and Marangoni propulsion is sufficient to trigger complex transition toward spontaneously reversing regimes in this minimal axisymmetric system, due to the strong nonlinearity introduced by the surfactant's advection by the chemically-driven flow field.

The full characterization of this oscillating regime is beyond the scope of the present analysis. Yet some preliminary features can be identified. In Figure~\ref{chaos}b, we use autocorrelation function to demonstrate that the typical duration of a single self-propulsion spurt is ${\lesssim 100}$ time units. At long times erratic oscillations cancel out and the droplet transport is due to a small drift with average dimensionless velocity $\sim 10^{-4}$. To illustrate the presence of the drift in chaotic oscillations observed at ${\pe=80}$, ${m=2}$, and ${\eta=1}$, we plot droplets mean square displacement (MSD) in Fig.~\ref{chaos}c and show that ${\text{MSD} \sim t^2}$ at long times. Yet, at this stage, it is unclear whether this identifies a persistent and preferred direction of motion in this regime or whether this slow drift is simply due to lthe excitation of low frequency subharmonics.
Discriminating these two effects and a full characterization of this chaotic regime requires much longer computations and is left for future research.

\section{Discussion}
\label{discussion}
The goal of this work is to elucidate how complex dynamical behavior, such as steady or chaotic propulsion, arises in individual chemically active drops. To this end, we proposed a minimal axisymmetric model of a solitary chemically active drop that stirs the flow in the bulk of surrounding surfactant solution due to a combined action of diffusiophoresis and the Marangoni effect. Our model allows for a fully-resolved description of the coupled hydrodynamic and advection-diffusion problems. We postulate that the drop features constant and isotropic chemical activity with a prescribed value of the flux of surfactant particles at its surface. The resulting droplet dynamics is characterized using both numerical simulations based on an azimuthal spectral decomposition of the concentration field and asymptotic analysis near the onset of self-propulsion. Surfactant advection by the surface-driven flows is the only nonlinear ingredient in this model, and is shown to be sufficient to enable not only the onset of self-propulsion from an isotropic steady state but also complex transitions between different dynamic behaviors, including steady-self-propulsion, stationary stirring of the flow and chaotic unsteady self-propulsion.

More specifically, our key results are as follows:
\begin{enumerate}[label=(\roman*)]

  \item
  Strong advection (e.g., large droplet size) may destabilize a steadily self-propelling drop. In this case, the droplet spontaneously stops and a symmetric extensile flow emerges, as shown in Figs.~\ref{regimes}a and~\ref{flows1}c. If advection is strengthened even further (i.e., increasing $\pe$), the symmetric state loses its stability and the droplet enters chaotic oscillations illustrated in Fig.~\ref{chaos}, characterized by random reversal of the direction of propulsion and short excursions of the velocity magnitude. This transition from steady self-propulsion to chaotic motion when $\pe$ is increased is reminiscent of the experimental observations of Ref.~\cite{Suga18} for the successive behavior of a gradually dissolving droplet, at least in the framework of the axisymmetric assumption of our approach.

  \item
  The thresholds corresponding to transitions between the dynamical regimes depend on the balance between diffusiophoresis and the Marangoni effect, quantified by the mobility contrast $m$. More specifically, these thresholds are observed to decrease (and saturate) with increasing $m$. Within the considered range of $\pe$, flow symmetrisation and chaotic oscillations are only observed for a large enough mobility ratio $m$: when diffusiophoresis is weak, a large value of $\pe$ is required for such complex dynamical states to develop. These results suggest that chaotic oscillations may not arise for purely Marangoni propulsion ($m\ll 1$) and that a small amount of diffusiophoretic behavior is needed. Yet, to confirm these results numerical simulations using a different approach might be needed as the spectral azimuthal expansion of the concentration converges slowly with the number of Legendre modes for large $\pe$, rendering the present approach prohibitively \blu{expensive} computationally.
  
  \item
  Linear stability analysis reveals that diffusiophoresis promotes the onset of higher-order modes of instability, as shown in Fig.~\ref{thresholds}.

\end{enumerate}

We argue that the sensitivity of the droplet nonlinear dynamics to the mobility contrast is corroborated by the predictions of linear stability analysis. In particular, the effect captured in Fig.~\ref{thresholds} is echoed by Eqs.~\eqref{ain_sol}--\eqref{aon_sol} that link the Stokes flow with the concentration filed near the droplet interface. We note that in the pure Marangoni case, ${m=0}$, the transfer coefficients relating the concentration and hydrodynamic modes decay asymptotically ${A_{i,n}, A_{o,n}} \rightarrow 0$ as $n \rightarrow \infty$, while in the presence of diffusiophoresis (${m \neq 0}$) these transfer coefficients remain finite as $n \rightarrow \infty$. That is, Marangoni effect damps the onset of higher-order modes in the expansion~\eqref{c_num_exp}, thus hindering flow symmetrization and subsequent onset of chaos.

In contrast with experimental results, self-propulsion of the drop seems to always slow down drastically after the onset of chaos in the present model, a feature which may well be a by-product of the axisymmetric assumption.  In this case, only two opposite directions of self-propulsion are allowed; thus, to change the direction of motion the drop has to first stop and then reverse its course. In turn, motionless drop corresponds to the trivial solution of the problem~\eqref{base} and system dynamics must be slow in vicinity of this fixed point solution. In contrast, in experimental systems, the direction of motion is not restricted, so active drops observed experimentally can change the direction of their self-propulsion without stopping. Three-dimensional dynamics of the concentration field and reorientation of the drop within the entire angular space may therefore open the possibility for other dynamical regimes such as rotation and spiralling motion as observed in experiments.

In addition, by introducing the competing effect of diffusiophoresis and Marangoni forcing, we demonstrated that the detailed behavior of the droplet's surface in response to a concentration gradient may sharply modify the properties of the flow field within the droplet. For instance, we demonstrated that the direction of the flow within the drop depends on the mobility contrast as well. Specifically, for ${m > \eta / 2}$, the recirculation within the drop is reversed, compared to the pure Marangoni case. We note that such flow reversal is a typical feature of the drops propelled by phoretic effects~\cite{Yang18} and hypothesize that this feature may serve as an indicator in experiments to gather the information about the physical mechanisms enabling active droplets mobility, and might also be present for more complex interface properties (e.g., viscoelastic properties or liquid-crystal droplets).

We emphasized here that the axisymmetric assumption, at the heart of the modeling followed here for a single drop, imposes some significant restriction in the dynamical behavior and a natural, albeit challenging, extension of the present work resides in the analysis of the system's bifurcation when this assumption is relaxed.

\acknowledgements{
This project has received funding from the European Research Council (ERC) under the European Union's Horizon 2020 research and innovation programme (grant agreement No 714027 to SM).
}

\appendix

\section{Asymptotic analysis}
\label{asymptotic}
The problem formulated by the dimensionless form of the Eqs.~\eqref{eqs_flow}-\eqref{slip} and~\eqref{stress}--\eqref{far} allows for a trivial solution,
\begin{equation}
  \label{base}
  \psi_i = \psi_o = 0, \quad
  C = -\frac{1}{r},
\end{equation}
corresponding to a motionless drop with isotropic concentration distribution and no fluid motion. This steady isotropic state is known to become unstable for finite $\pe$ when $m=0$ and $m\gg 1$~\cite{Michelin13b,Izri14}, and it is therefore expected that this transition to self-propulsion is a generic feature for all $m$. 

In this section, the asymptotic analysis of the steady flows emerging near the base state~\eqref{base} is carried out, with the goal to elucidate the system dynamics near the onset of self-propulsion. Since the analysis follows the logic of our earlier work~\cite{Morozov18}, we keep the technical details to a minimum.

In the vicinity of the base state, the steady flow field is weak and the stream function can be expanded as follows,
\begin{equation}
  \label{psi_exp}
  \left( \psi_i, \psi_o \right) \left( r, \mu \right)
  = \epsilon \left( \psi_i^{(1)}, \psi_o^{(1)} \right)
  + \epsilon^2 \left( \psi_i^{(2)}, \psi_o^{(2)} \right)
  + \ldots
\end{equation}
where $\epsilon \ll 1$. Since the flow field is small, advection is weak and surfactant concentration distribution features a boundary layer at $r \rightarrow \infty$~\cite{Acrivos62}. Accordingly, we employ matched asymptotic expansion of the concentration field,
\begin{align}
  \label{c_exp}
  & C(r,\mu) = -\frac{1}{r} + \epsilon C^{(1)} + \epsilon^2 C^{(2)} + \ldots, \\
  & H(\rho,\mu) = \epsilon H^{(1)} + \epsilon^2 H^{(2)} + \ldots,
\end{align}
where $\rho\equiv r/\epsilon\sim 1$ ($r \gg 1$) and $H$ denotes the concentration of surfactant far from the drop and satisfies the rescaled advection/diffusion equation given by
\begin{multline}
  \label{c_far_eq}
  - \epsilon \pe \left(
        \frac{\partial \psi_o}{\partial \mu} \frac{\partial H}{\partial \rho}
      - \frac{\partial \psi_o}{\partial \rho} \frac{\partial H}{\partial \mu}
    \right)
  \\ = \frac{\partial}{\partial \rho} 
      \left( \rho^2 \frac{\partial H}{\partial \rho} \right) 
  + \frac{\partial}{\partial \mu} 
      \left( \left( 1 - \mu^2 \right) \frac{\partial H}{\partial \mu} \right).
\end{multline}

We now substitute expansions~\eqref{psi_exp}--\eqref{c_exp} into the dimensionless form of the Eqs.~\eqref{eqs_flow}-\eqref{slip} and~\eqref{stress}--\eqref{far} and in Eq.~\eqref{c_far_eq} and collect the terms at the same order of $\epsilon$, thus obtaining a sequence of linear problems. In the following we solve the first two problems in the sequence and extract the threshold of spontaneous self-propulsion as well as the self propulsion velocity.

\subsection{Problem at $\epsilon$}
\label{lin}
The first problem in the sequence, the problem at $\epsilon$, reads in the near field,
\begin{align}
  \label{lin_ad}
  & \nabla^2 C^{(1)} 
    = -\frac{\pe}{r^4} \frac{\partial \psi_o^{(1)}}{\partial \mu}, \\
  \label{lin_bc1}
  & \frac{\partial C^{(1)}}{\partial r} = 1, \quad
  \frac{\partial \psi_i^{(1)}}{\partial \mu} 
  = \frac{\partial \psi_o^{(1)}}{\partial \mu} = 0, \\
  & \frac{\partial \psi_i^{(1)}}{\partial r} 
  - \frac{\partial \psi_o^{(1)}}{\partial r}
  = \frac{2 + 3 \eta}{\eta \left( 3 + 1/m \right)} \left( 1 - \mu^2 \right)
    \frac{\partial C^{(1)}}{\partial \mu}, \\
  \label{lin_stress}
  & \left( \frac{\partial^2}{\partial r^2} - 2 \frac{\partial}{\partial r}
    - \left( 1 - \mu^2 \right) \frac{\partial^2}{\partial \mu^2} \right) 
        \left( \psi_o^{(1)} - \eta \psi_i^{(1)} \right) \nonumber \\
  & \qquad \qquad = \frac{2 + 3 \eta}{1 + 3 m} \left( 1 - \mu^2 \right) 
      \frac{\partial C^{(1)}}{\partial \mu}
  \quad \text{at  } r = 1,
\end{align}
and in the far field,
\begin{align}
  \label{lin_ad_far}
  &2 \pe \, a_{o,1}^{(1)} \left( \mu \frac{\partial H^{(1)}}{\partial \rho} 
    + \frac{1 - \mu^2}{\rho} \frac{\partial H^{(1)}}{\partial \mu} \right)
    \nonumber \\
    &+ \frac{1}{\rho^2} \left[ 
        \frac{\partial}{\partial \rho} 
          \left( \rho^2 \frac{\partial H}{\partial \rho} \right) 
      + \frac{\partial}{\partial \mu} \left( \left( 1 - \mu^2 \right) 
          \frac{\partial H}{\partial \mu} \right)
    \right]
  = 0, \\
  \label{lin_ad_far_bc}
  &\textrm{with    }H \rightarrow 0 \qquad \text{as  } \rho \rightarrow \infty.
\end{align}
Naturally, collecting the terms at $\epsilon$ is equivalent to linearization of the problem near the base state~\eqref{base}. Recall that we consider steady flows, so the linearized problem at hand yields a set of neutrally stable eigenmodes of the droplet.

Following Ref.~\cite{Morozov18} we assume the solution of Eqs.~\eqref{lin_ad} and~\eqref{lin_ad_far} in the form,
\begin{align}
  \label{gen_sol}
  & C^{(1)}( r, \mu ) = \sum\limits_{n=0}^\infty C_n^{(1)} (r) L_n(\mu), \\
  & H^{(1)}( \rho, \mu ) 
    = \frac{ e^{-\rho_s \mu} }{ \sqrt{ \left| \rho_s \right| } }
      \sum\limits_{n=0}^\infty h_n^{(1)} 
        K_{n+1/2} \left( \left| \rho_s \right| \right) L_n(\mu),
\end{align}
and find the following expressions for $C_n^{(1)}(r)$,
\begin{align}
  \label{c1_gen_sol1}
  & C_0^{(1)}(r) = \frac{c_0^{(1)}}{r} + d_0^{(1)}, \\
  & C_1^{(1)}(r) = \frac{c_1^{(1)}}{r^2} + d_1^{(1)} r 
    + \pe a_{o,1}^{(1)} \frac{ 1 + 2 r^3}{2 r^3}, \\
  \label{c1_gen_sol2}
  & C_n^{(1)}(r) \big|_{n > 1} = \frac{c_n^{(1)}}{r^{n+1}} + d_n^{(1)} r^n
    + \pe a_{o,n}^{(1)} \frac{n + (n + 1) r^2}{2 r^{n+2}}.
\end{align}

We employ Van Dyke's matching rule~\citep{Holmes95}, to match~$C^{(1)}$ and $H^{(1)}$ in the region ${\epsilon \ll \rho \ll 1}$ and then substitute $\psi_i^{(1)}$, $\psi_o^{(1)}$, and $C^{(1)}$ given by~\eqref{innerflow}, \eqref{outerflow}, and~\eqref{gen_sol}--\eqref{c1_gen_sol2}, respectively, into the boundary conditions~\eqref{lin_bc1}--\eqref{lin_stress}. Since we consider the case of a steady flow, projection of the result onto the $n$-th Legendre polynomial yields a sequence of sets of homogeneous linear algebraic equations for the constant amplitudes $a_{i,n}^{(1)}$, $a_{o,n}^{(1)}$, and $c_n^{(1)}$. Solvability condition of the $n$-th set of equations reads,
\begin{equation}
  \label{pe_crit_def}
  \pe = \pe_n \equiv \begin{cases}
    4, & n = 1 \\
    \\
    \dfrac{4 \left( n + 1 \right)
      \left( 1 + \eta \right) \left( 1 + 3 m \right)}
    {\left( 2 + 3 \eta \right) \left( \left[ 2 n + 1 \right]^{-1} + m \right)}, 
    & n > 1
  \end{cases}.
\end{equation}

\begin{figure}
\centering
  \includegraphics[scale=0.6]{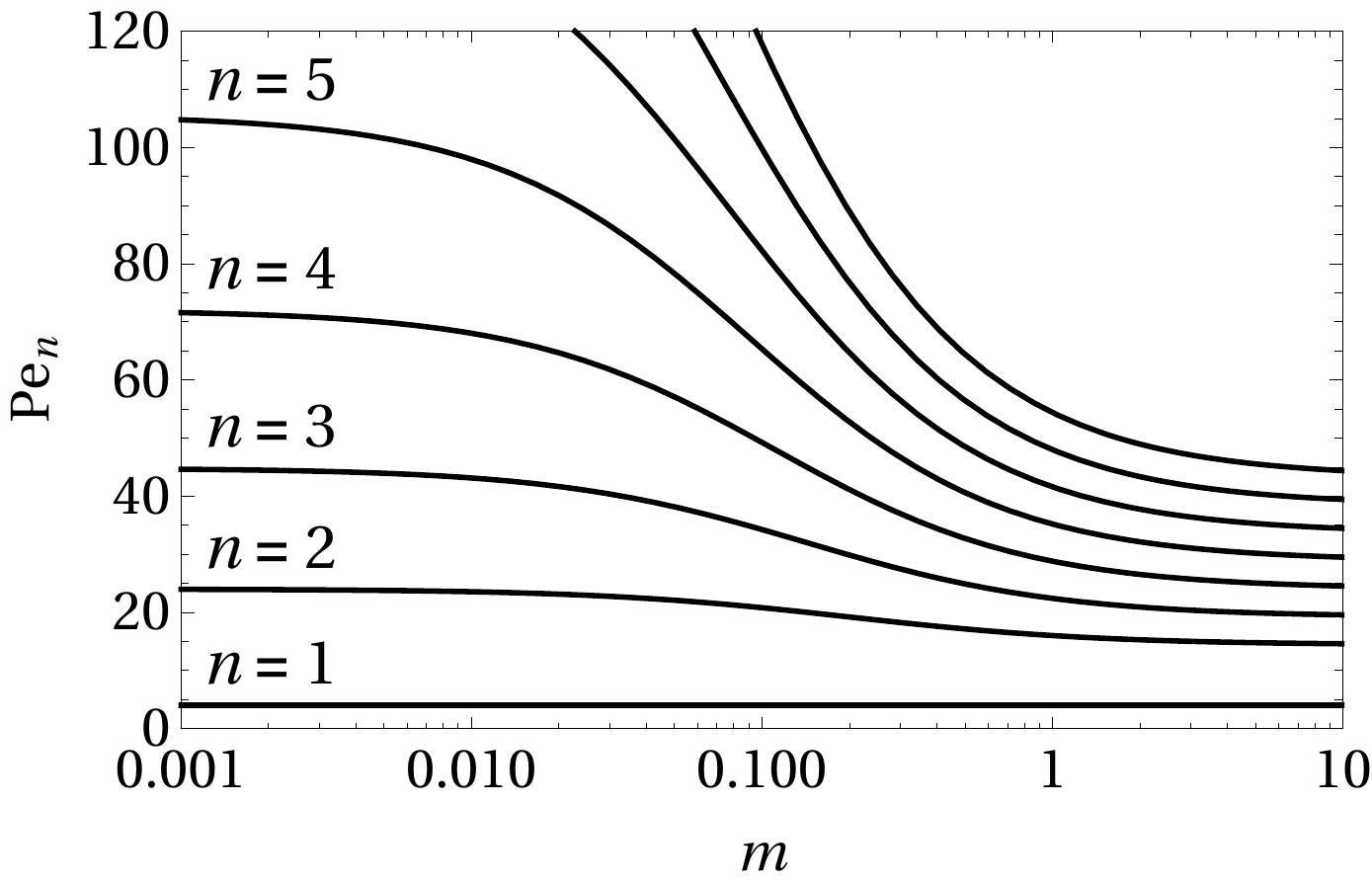}
  \caption{
    Thresholds of the first eight modes of monotonic instability.
  }
  \label{thresholds}
\end{figure}

In essence, condition~\eqref{pe_crit_def} establishes that $n$-th neutrally stable eigenmode of the linearized problem exists at a distinct point $\pe = \pe_n$. As a result, in vicinity of the point $\pe = \pe_n$, only the $n$-th eigenmode may be excited near the base state~\eqref{base} and, thus, $\pe_n$ represents the threshold of the $n$-th mode of monotonic instability. Thresholds of the first eight instability modes are shown in Fig.~\ref{thresholds}. Recall that $m = 0$ corresponds to Marangoni-dominated flow, whereas diffusiophoresis prevails in the limit of $m \rightarrow \infty$. It is easy to see that, although the threshold of the first mode, $\pe_1 = 4$ remains constant, diffusiophoresis promotes the onset of higher instability modes which is crucial for the droplet dynamics away from the threshold.

\subsection{Problem at $\epsilon^2$}
We now aim to obtain the terminal velocity of the drop near the instability threshold. To this end we focus on the mode with $n=1$ (i.e., the only mode featuring nonzero velocity as $r \rightarrow \infty$) given by,
\begin{align}
  \label{o1_sol_n1}
  & a_{i,0}^{(1)} = 3 A_1 \frac{\eta - 2 m}{2 \eta ( 1 + 3 m )}, \quad
  c_0^{(1)} = 0, \\
  & d_0^{(1)} = \pe_1 A_1, \quad
  c_1^{(1)} = -\frac{3 \pe_1 A_1}{4}, \quad
  d_1^{(1)} = 0,
\end{align}
where $A_1$ is and unknown constant. Following Ref.~\cite{Morozov18}, we assume that the P{\'e}clet number is close to $\pe_1$, namely,
\begin{equation}
  \pe = \pe_1 + \epsilon \delta,
\end{equation}
and obtain the solvability condition of the problem at $\epsilon^2$,
\begin{equation}
  A_1 \left( A_1 - \delta / 32 \right) = 0.
\end{equation}
And conclude that in vicinity of $\pe_1$ the droplet is either quiescent (${A = 0}$ or self-propels with the terminal velocity ${U_\infty = \delta / 16}$. Note that due to the choice of dimensionless velocity, dimensionless $U_\infty$ does not depend on $m$.

\bibliographystyle{unsrt}
\bibliography{nonlin}

\end{document}